\documentclass[12pt]{article}

\usepackage{epsfig}

\def\lsim{\raise0.3ex\hbox{$<$\kern-0.75em\raise-1.1ex\hbox{$\sim$}}}
\def\gsim{\raise0.3ex\hbox{$>$\kern-0.75em\raise-1.1ex\hbox{$\sim$}}}
\def\Chi{\raise0.4ex\hbox{$\chi$}}
\def\Tr{{\rm tr}}

\newcommand{\beqn} {\begin{equation}}
\newcommand{\eqn} {\end{equation}}
\newcommand{\refe}[1]{(\ref{#1})}

\begin{document}
\thispagestyle{empty}
\def\thefootnote{\fnsymbol{footnote}}
\setcounter{footnote}{1}
\hfill
   \begin{minipage}[t]{4cm}
   \begin{flushright}
    BI-TP 99/17\\
   \end{flushright}
   \end{minipage}
\vspace*{2cm}
\begin{center}
{\Large\bf Heavy Quark Potentials in Quenched QCD at High Temperature}   \\
\end{center}
\vspace*{0.3cm}
\begin{center}
Olaf Kaczmarek$^{\rm a}$, Frithjof Karsch$^{\rm a,b}$,
Edwin Laermann$^{\rm a,c}$ and Martin L\"utgemeier$^{\rm a}$
\\[.3cm]
{\it $^{\rm a}$ Fakult\"at f\"ur Physik, Universit\"at Bielefeld,
D-33615 Bielefeld, Germany}\\
{\it $^{\rm b}$ Center for Computational Physics, University of Tsukuba,
Ibaraki-305, Japan}\\
{\it $^{\rm c}$ National Institute for Nuclear Physics and High Energy Physics
(NIKHEF), \\ Postbus 41882, 1009 DB Amsterdam, The Netherlands}
\end{center}
\vspace*{2.5cm}
{\large\bf Abstract}

\noindent
Heavy quark potentials are investigated at high temperatures.
The temperature range covered by the analysis extends from
$T$ values just below the deconfinement temperature
up to about $4 T_c$ in the deconfined phase. We simulated
the pure gauge sector of QCD on lattices with temporal
extents of 4, 6 and 8 with spatial volumes of $32^3$.
On the smallest lattice a tree level improved action was
employed while in the other two cases the standard Wilson
action was used. Below $T_c$ we find a temperature dependent
logarithmic term contributing to the confinement potential
and observe a string tension which decreases with rising
temperature but retains a finite value at the deconfinement
transition. Above $T_c$ the potential is Debye-screened,
however simple perturbative predictions do not apply.

\vspace{1.5cm}
PACS-Indices : 5.70.Ce 12.38.Gc 12.39.Pn


\def\thefootnote{\arabic{footnote}}
\setcounter{footnote}{0}
\clearpage

\section{Introduction}
\label{intro}

The static quark potential at high temperatures is
interesting for several reasons.
Phenomenologically, the properties of quark
bound states, in particular of heavy quarkonia,
can be derived from potential models quite
successfully. It is then important to compute
the temperature dependence of the potential
as this might lead to observable consequences
in heavy ion collision experiments.
Notably, it has been suggested to use the suppression
of $J/\Psi$ and $\Psi'$ production \cite{Matsui} as a signal for
the quark-gluon plasma. For this purpose, a detailed
knowledge of the temperature dependence of the potential
appears very helpful \cite{Mehr}.

Moreover, it is well known that a linearly increasing potential
at large distances arises naturally from a string picture
of confinement. As long as one stays in the confined phase of QCD,
string models then also predict a definite
behaviour of the potential at finite temperatures
\cite{Alvarez,Schierholz,Gao}.
These predictions ought to be tested by lattice analyses.

In close vicinity of the deconfinement transition temperature
the static quark potential and the mass gap i.e. the potential
integrated over perpendicular directions are sensitive
to the order of the phase transition.
In colour $SU(3)$ the observation of a finite mass gap
at the critical temperature supported
that the transition is of first order \cite{Ukawa},
while in $SU(2)$
a continuous decrease to zero with the appropriate Ising
critical exponents was found \cite{Juergen}.

In the deconfined phase, asymptotic freedom suggests that
at high temperatures
the plasma consists of weakly interacting quarks and gluons.
Previous numerical studies
[8-12]
have, however, shown that
non-perturbative phenomena prevail up to temperatures of at least
several times the critical temperature. In particular, the
heavy quark potential did not show the simple Debye-screened
behaviour anticipated from a resummed lowest-order perturbative
treatment \cite{Nadkarni1}. This might not be too surprising
as various non-perturbative modes may play a role in the
long distance sector of the plasma \cite{Janos}.
It is then important to quantify colour screening effects
by a genuinely non-perturbative approach.

In the present paper we compute the static quark potential
in the pure gluonic sector of QCD. We investigate the
temperature dependence of the potential over a range of
temperatures from 0.8 to about 4 times the critical temperature
$T_c$. The analysis is based on gluon configurations generated
on lattices of size $32^3 \times N_\tau$ with $N_\tau = 4,6$ and $8$.
This enables us to gain some control over finite lattice spacing
artefacts.
On the smallest lattice a tree level improved gauge action was used
while on the two bigger lattices a standard Wilson action
was employed.
We go beyond previous studies of the potential in so far the
temperature range is covered more densely and also because
a large set of lattice distances was probed. This
helps to extract fit parameters with higher reliability.

The paper is organized such that the next section summarizes
theoretical expectations on the behaviour of the potential
both below and above the transition temperature. In
section~\ref{below}
we present and discuss our results for the potential in the
confined phase. Section~\ref{above} contains our findings for
temperatures above $T_c$ and section~\ref{conclusion} the
conclusion.

\section{Theoretical Expectations}
\label{theory}

Throughout this paper the potential is
computed from Polyakov loop correlations
\beqn
\langle L({\vec 0}) L^\dagger ({\vec R}) \rangle =
\exp\{ - V(|{\vec R}|,T)/T\}
\label{corrfct}
\eqn
where
\beqn
L({\vec x})= \frac{1}{3} \Tr \prod_{\tau = 1}^{N_\tau}
U_0({\vec x},\tau)
\eqn
denotes the Polyakov loop at spatial coordinates ${\vec x}$.
In the limit $R \rightarrow \infty$ the correlation function
should approach the cluster value
$|\langle L(0)\rangle |^2$ which vanishes if the potential
is rising to infinity at large distances (confinement)
and which acquires a finite value in the deconfined phase.

In the limit where the flux tube between two static quarks can
be considered as a string, predictions about the
behaviour of the potential are available from computations
of the leading terms arising in string models. For zero
temperature one expects
\beqn
V(R) = V_0 -\frac{\pi}{12} \frac{1}{R} + \sigma R
\label{zero}
\eqn
where $V_0$ denotes the self energy of the quark lines, $\sigma$
is the string tension and the Coulomb-like $1/R$ term stems
from fluctuations of the string \cite{Luescher}.
Eq.~(\ref{zero}) generally gives a good description of
the zero temperature ground-state potential although it has
been shown \cite{Kuti} that the excitation spectrum meets string
model predictions only at large quark pair separations.
For non-vanishing temperatures
below the critical temperature of the transition to deconfinement,
a temperature-dependent potential has been computed \cite{Gao} as
\begin{eqnarray}
V(R,T) & = & V_0 - \left[ \frac{\pi}{12} - \frac{1}{6} {\rm arctan}(2RT)\right]
\frac{1}{R}  \nonumber \\
& & \mbox{}+\left[ \sigma - \frac{\pi}{3} T^2 + \frac{2}{3}T^2
{\rm arctan}(\frac{1}{2RT})\right] R
+ \frac{T}{2} \ln(1+(2RT)^2)
\label{Gao}
\end{eqnarray}
In the limit $R \gg 1/T$ this goes over into
\beqn
V(R,T) = V_0 + \left[\sigma -\frac{\pi}{3}T^2 \right] R
+ T \ln(2RT)
\label{gerrit}
\eqn
which had been calculated previously \cite{Schierholz}.
Note the logarithmic term which originates from transverse
momentum fluctuations\footnote{In the context of analyzing
numerical data this term has been mentioned
in \cite{Ukawa,Kanaya} and was discussed
in detail in \cite{Juergen}.}.
So far, it has been left open whether the string tension $\sigma$
appearing in eqs.~\refe{Gao} and \refe{gerrit} is identical to the
zero temperature value.
In the context of a low temperature or large $R$ expansion, the
temperature dependent terms appearing in eqs.~\refe{Gao} and \refe{gerrit}
should, however, be considered as thermal corrections to the zero
temperature string tension.
An explicitly temperature-dependent string tension was computed
by means of a $1/D$ expansion \cite{Alvarez}
\beqn
\frac{\sigma(T)}{\sigma(0)} = \sqrt{1-\frac{T^2}{T_c^2}}
\label{eq:alvarez}
\eqn
where $T_c$ was obtained as
\beqn
T_c^2 = \frac{3}{\pi(D-2)} \sigma(0) \quad .
\eqn
Note, however, that for $D \rightarrow \infty$ the phase transition
is of second order leading to a continuous vanishing
of the string tension at the deconfinement temperature.
In colour $SU(2)$, which also exhibits a second order transition,
it was established \cite{Juergen}
that $\sigma(T)$ vanishes $\sim (\beta_c-\beta)^\nu$ with
a critical exponent $\nu$ taking its 3-D Ising value
of 0.63 as suggested by universality. In the present case of
colour $SU(3)$ one expects a discontinuous behaviour
and a non-vanishing string tension
at the critical temperature.

In the deconfined phase the Polyakov loop acquires a
non-zero
value. Thus, we can normalize the correlation function
to the cluster value $| \langle L \rangle |^2$, thereby removing
the quark-line self energy contributions.
Moreover, the quark-antiquark pair can be in either a colour singlet or
a colour octet state. Since in the plasma phase quarks are deconfined
the octet contribution does not vanish\footnote{It is, however,
small compared to the singlet part. This is true perturbatively,
see eq.~\refe{octet}, as well as numerically \cite{Attig}.}
and the Polyakov loop correlation is
a colour-averaged mixture of both
\beqn
e^{-V(R,T)/T} = \frac{1}{9}e^{-V_1(R,T)/T} + \frac{8}{9}
e^{-V_8(R,T)/T}
\eqn
At high temperatures, perturbation theory predicts \cite{Nadkarni1}
that $V_1$ and $V_8$ are related as
\beqn
V_1 = - 8 V_8 + {\cal O}(g^4)
\label{octet}
\eqn
Correspondingly, the colour-averaged potential is given by
\beqn
\frac{V(R,T)}{T} = -\frac{1}{16} \frac{V_1^2(R,T)}{T^2}
\label{average_pot}
\eqn
Due to the interaction with the heat bath the gluon acquires
a chromo-electric mass $m_e(T)$ as the IR limit of the
vacuum polarization tensor. To lowest order in
perturbation theory, this is obtained as
\beqn
\left(\frac{m_e^{(0)}(T)}{T}\right)^2 = g^2(T)\left(\frac{N_c}{3}
+\frac{N_F}{6}\right)
\label{pertmu}
\eqn
where $g(T)$ denotes the temperature-dependent
renormalized coupling, $N_c$ is the number of colours and $N_F$ the
number of quark flavours.
The electric mass is also known in next-to-leading order \cite{rebhan}
in which it depends on an anticipated chromo-magnetic
gluon mass although the magnetic gluon mass itself
cannot be calculated perturbatively.
Fourier transformation of the gluon propagator
leads to the Debye-screened
Coulomb potential for the singlet channel
\beqn
V_1(R,T) = - \frac{\alpha(T)}{R} e^{-m_e(T)R}
\label{singlet_pot}
\eqn
where $\alpha(T) = g^2(T) (N_c^2-1)/(8\pi N_c)$ is the renormalized
T-dependent fine structure constant.
It has been stressed \cite{Gale} that eq.~\refe{singlet_pot}
holds only in the IR limit $R \rightarrow \infty$ because momentum
dependent contributions to the vacuum polarization tensor have
been neglected.
Moreover, at temperatures just
above $T_c$ perturbative arguments will not apply
so that we have chosen to attempt a parametrization of the numerical
data with the more general ansatz \cite{Attig}
\beqn
\frac{V(R,T)}{T} = - \frac{e(T)}{(RT)^d} e^{-\mu(T)R}
\label{free_d}
\eqn
with an arbitrary power $d$ of the $1/R$ term,
an arbitrary coefficient $e(T)$ and a simple
exponential decay determined by a general screening mass $\mu(T)$.
Only for $T \gg T_c$ and large distances
we expect that $d \rightarrow 2$ and $\mu(T) \rightarrow 2 m_e(T)$,
eq.~\refe{average_pot},
corresponding to two-gluon exchange.

\section{Results below $T_c$}
\label{below}

The results to be presented here as well as in the next section
are based on two different sets of data.
The first set, referred to as (I) in the following,
was generated with a tree-level Symanzik-improved
gauge action consisting of $1\times 1$ and $2 \times 1$ loops.
The lattice size was $32^3 \times 4$. We used a pseudo-heatbath
algorithm \cite{CM} with FHKP updating \cite{FHKP} in the $SU(2)$
subgroups. Each heatbath iteration is supplemented by 4
overrelaxation steps \cite{Adler}. To improve the signal in calculations
of Polyakov loop correlation functions
link integration \cite{linkintegral} was employed.
For each $\beta$-value the data set consists of 20000 to 30000 measurements
separated by one sweep.

The second set of data (II) was obtained as a by-product of
earlier work,
the analysis of the equation of state \cite{EOS}. The
gauge configurations used in the present study were
generated with the standard Wilson gauge action on lattices
of size $32^3 \times 6$ and $32^3 \times 8$.
The same algorithm as for (I) was employed.
The statistics amounts to 1000 to 4000 measurements
separated by 10 sweeps for the $N_\tau=6$ data and
between 15000 and 30000 measurements at each sweep
for $N_\tau =8$.
The errors on the potentials as well as the fit parameters
were determined by
jackknife in both cases.

\begin{figure}[htb]
\epsfig{file=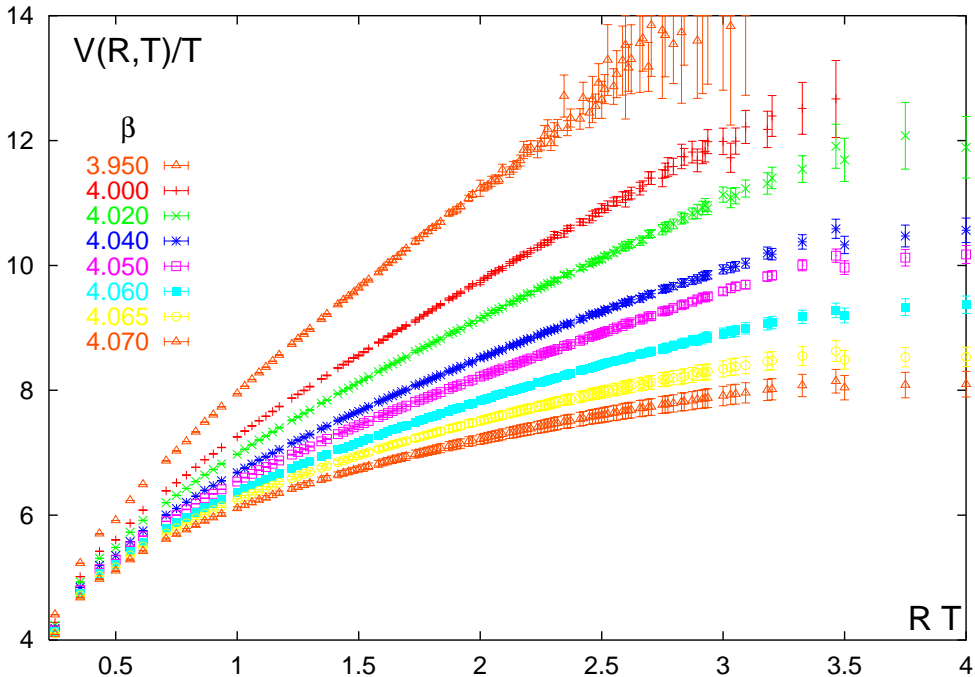,width=134mm}
\caption{The potentials in units set by the temperature
         at the $\beta$ values analyzed for $N_\tau=4$
         (case I). The critical coupling for this action
         on our lattice size has been determined as
         $\beta_c = 4.0729(3)$ \cite{BKP}.
         }
\label{fig:nt4}
\end{figure}

\begin{figure}[htb]
\epsfig{file=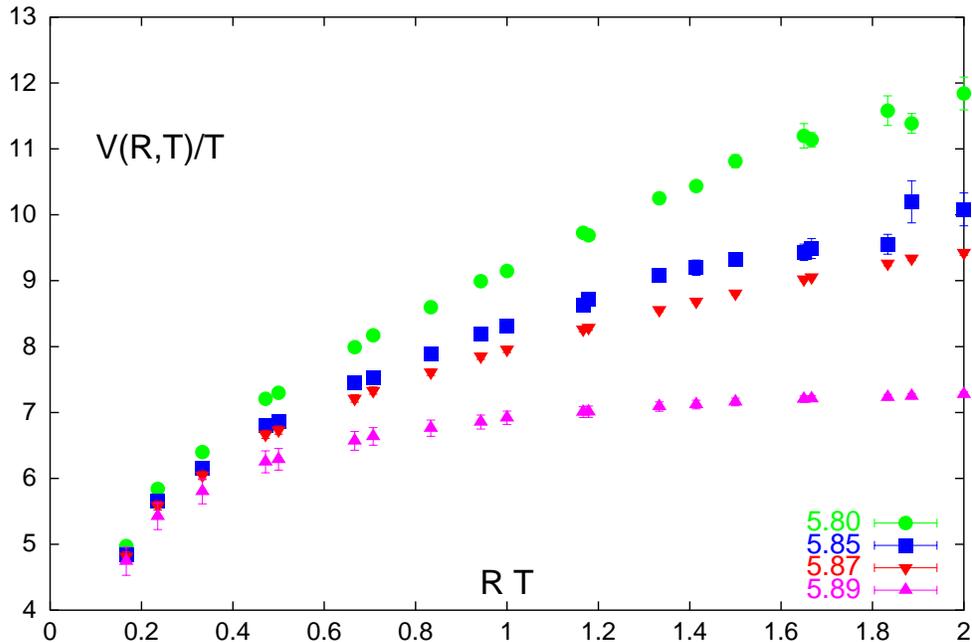,width=134mm}
\vspace{-4mm}
\caption{The same as figure \ref{fig:nt4} except for
         $N_\tau=6$ (case II). The critical coupling
         is $\beta_c = 5.8938(11)$ \cite{EOS}.
         }
\label{fig:nt6}
\end{figure}

\begin{figure}[htb]
\epsfig{file=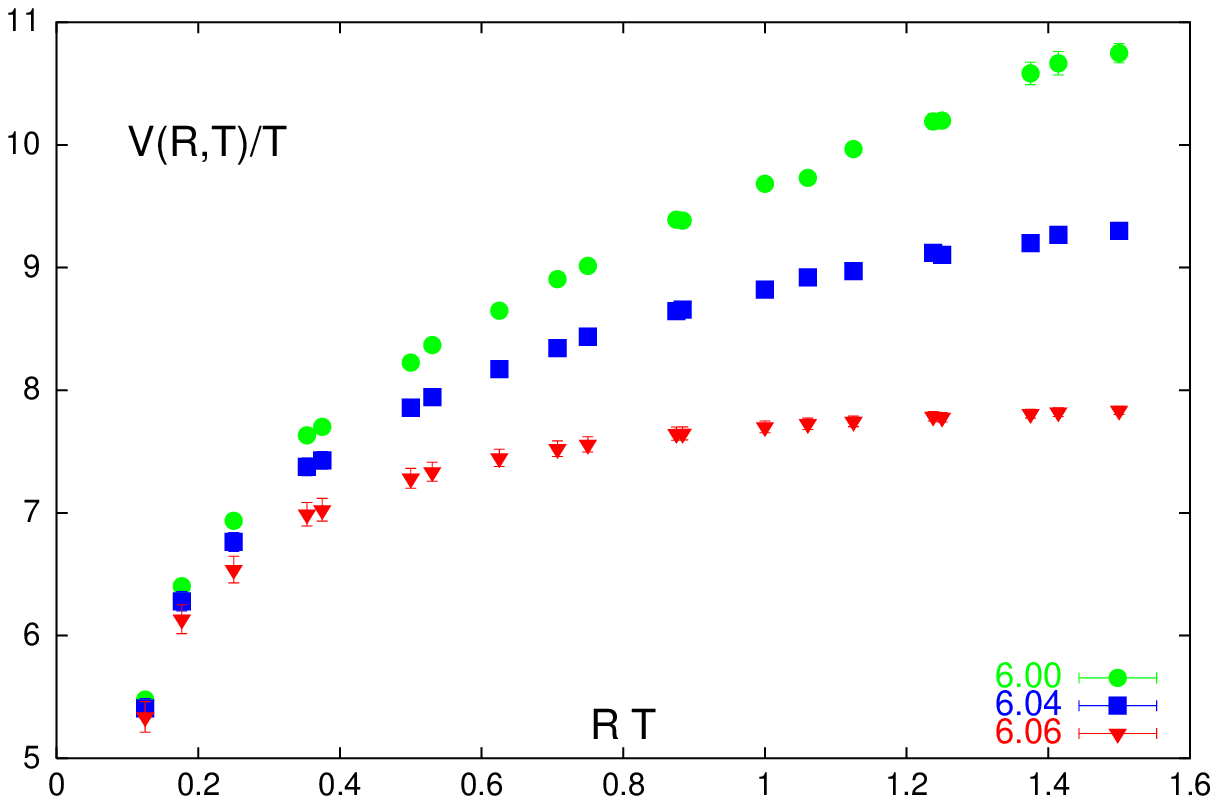,width=134mm}
\vspace{-4mm}
\caption{The same as figure \ref{fig:nt4} except for
         $N_\tau=8$ (case II).  The critical coupling
         is $\beta_c = 6.0609(9)$ \cite{EOS}.
         }
\label{fig:nt8}
\end{figure}

The lattice results for the potential at temperatures
below $T_c$ are shown in Figures
\ref{fig:nt4},~\ref{fig:nt6} and \ref{fig:nt8}. The correlation functions,
eq.~\refe{corrfct},
have been computed not only for on-axis separations but
also for some, in the case I almost all, off-axis
distance vectors $\vec{R}$. Although the lattice spacing
for the $N_\tau=4$ data is larger than for the other two
lattice sizes rotational symmetry is quite well satisfied
due to the use of an improved action in this case.
As we will focus on the intermediate to large distance behaviour of
the potential, it was not attempted to specifically treat the
deviations from rotational invariance at small separations.
Note that the
distances covered by the data extend to $RT \lsim 4$
for (I) while in case II we could obtain signals up to
$RT \,\lsim \,2$.

The potentials have first been fitted to eq.~\refe{Gao} with two
free parameters, the self-energy $V_0$ and a possibly
temperature-dependent string tension $\sigma(T)$.
These fits work rather well even when data at small separations
are included because the fit ansatz also accounts for a $1/R$ piece
in the potential. The results to be quoted for the string tension,
Table 1,
have, however, been obtained
when the data at small separations
are excluded from the fit. Typically, a minimal distance
of $RT \simeq 1/2$ was chosen. The fits are stable under
variation of $R_{\rm min}$ in this ballpark and return good $\Chi^2$ values.
Varying the maximum distance to be fitted does not lead to
noticeable changes of the results.
This holds for all three lattices.

\begin{figure}[htb]
\epsfig{file=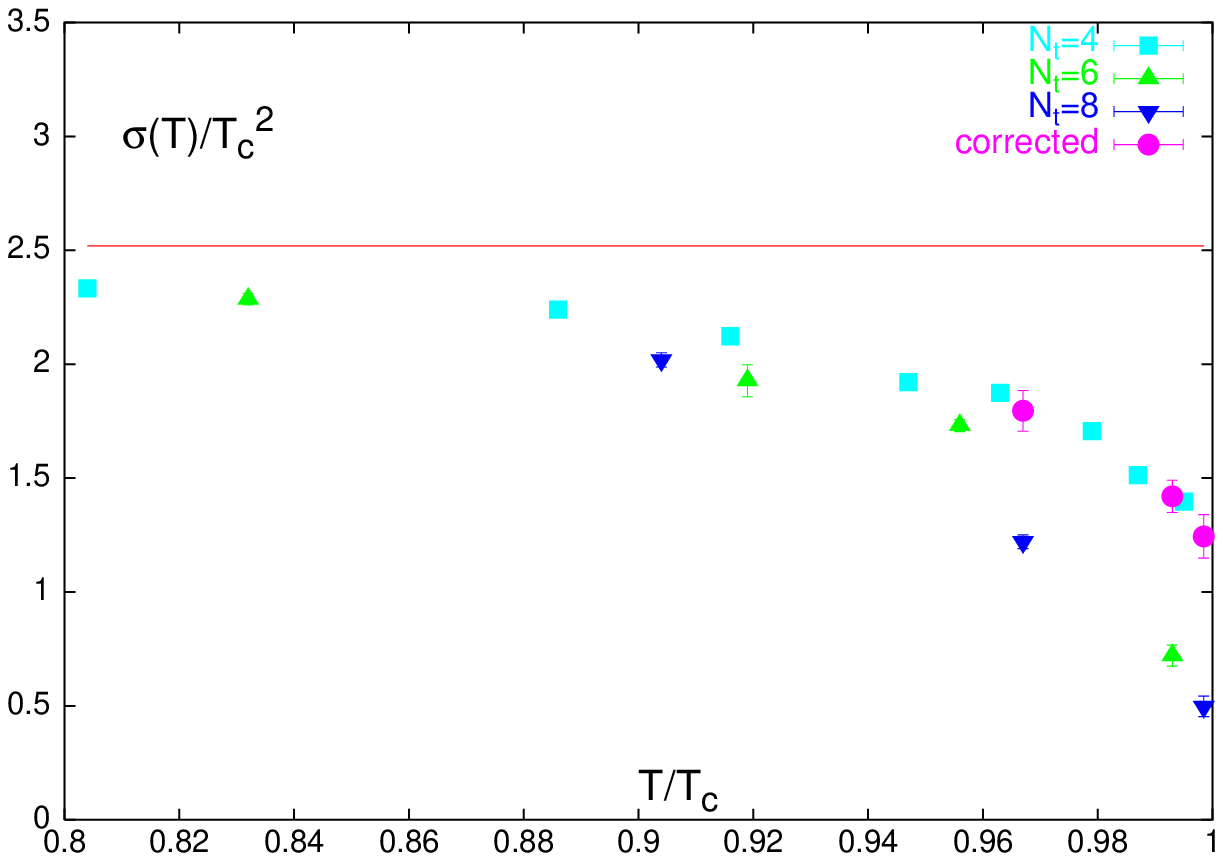,width=154mm}
\caption{The string tension at non-vanishing temperature
         as obtained from fits with
         eq.~\refe{Gao}. The line denotes the zero
         temperature string tension. In both cases
         the string tension was normalized to $T_c$.
         For an explanation of the `corrected' data points
         see text.
         }
\label{fig:gao}
\end{figure}

The results for the string tension, normalized to the
critical temperature squared, $\sigma/T_c^2$,
are summarized in Figure~\ref{fig:gao}.
The temperature scale has been determined from measurements
of the string tension at $T=0$ \cite{EOS,Tim,Beinlich}.
The finite temperature string tension is compared to these
results at zero temperature,
$\sigma(0)/T_c^2$, shown as the line in the figure.
Quite clearly, in the investigated temperature range
there are substantial deviations
from the zero temperature string tension.
These deviations amount to about 10 \% at $T/T_c = 0.8$
and become larger when the temperature is raised.

Close to $T_c$ the results from
$N_\tau = 6$ and $8$ at first sight do not seem to agree with
the numbers coming from the $N_\tau = 4$ lattices. However,
recall that the $SU(3)$ quenched theory exhibits a first order
transition with the coexistence of hadron and plasma phase at
the critical
temperature. The tunneling rate between the two
phases decreases exponentially
$\sim \exp(-2 \hat\sigma \times (N_\sigma/N_\tau)^2)$
where $\hat\sigma = \sigma_I/T_c^3$ is the normalized
interface tension.
The lattices of data set (II) have a smaller aspect ratio
of $N_\sigma / N_\tau= 4$ and 5.33, respectively, than the
$N_\tau = 4$ lattice whose aspect ratio is 8.
Correspondingly,
the ensemble of configurations of the second set contains
(more) configurations in the ``wrong'', the deconfined phase.
In fact, close
to $T_c$, Polyakov loop histograms reveal this two-state
distribution for $N_\tau = 6$ and 8, with a clear separability
between the two Gaussian-like peaks.
Such a two-state signal is absent for the $N_\tau = 4$
data. Carrying out the averaging of the potential only
over configurations with Polyakov loops in the confined peak
leads to the corrected data points in Figure~\ref{fig:gao}.
At temperatures not so close
to $T_c$ this separation of phases is not possible anymore
as the Polyakov loop histogram has tails into the deconfined phase
but it is not clear where one should set the cut.

When we apply this correction the agreement of the results from
the three different lattices is evident. This shows that the
temperature dependence of the string tension is not subject to
severe discretisation effects.
Moreover, the functional form of the fit ansatz eq.~\refe{Gao},
as suggested by string
model calculations, describes the behaviour of the lattice data
quite well.
However, with increasing temperature we observe a
substantial decrease of the string tension away from its zero
temperature value.
Since the fit ansatz, eq.~\refe{Gao}, contains already a
$\pi/3 \ast T^2$ term, the decreasing slope of the linear
part of the potential can not solely be accounted for by
this leading correction.

In order to analyze the linear rising part of the potential in
a more model-independent way,
in a second round of fits we
have compared our data with the ansatz
\beqn
V(R,T) = V_0 + \sigma(T) R + C T \ln (2 R T)
\label{gerrit2}
\eqn
Note that this ansatz differs from eq.~\refe{gerrit}
in so far it summarizes all the linear dependence on the
distance $R$ by an explicitly temperature dependent
string tension $\sigma(T)$. Due to its lacking of a
$1/R$ piece this formula is very well capable of describing
the data
but only if the fit is applied to large distances of $R T \geq 1$.
For data set (II) this requirement leaves not too many data points
to be fitted. In this case we checked that eq.~\refe{gerrit2}
is able to parametrize the potential. However, since we do not have
as much room to check for stability of the results as one would
wish, we refrain from quoting results for data set (II). In case I
we do have enough distances and obtain fits with good $\Chi^2$ values
which are stable under variation of the minimal distance to be
included in the minimization.

On data set (I) we clearly observe the logarithmic term
contained in eqs.~\refe{gerrit} and \refe{gerrit2}.
The fits return values for
the coefficient $C$ of the logarithm which
are equal to 1 within an error margin of less than 10 \%.
We thus confirm a logarithmic piece in the potential with a
strength as anticipated from the string model
calculation \cite{Schierholz} or, equivalently,
a subleading power-like $1/R$-factor with power 1 contributing
to the Polyakov-loop correlation function.

\begin{figure}[htb]
\epsfig{file=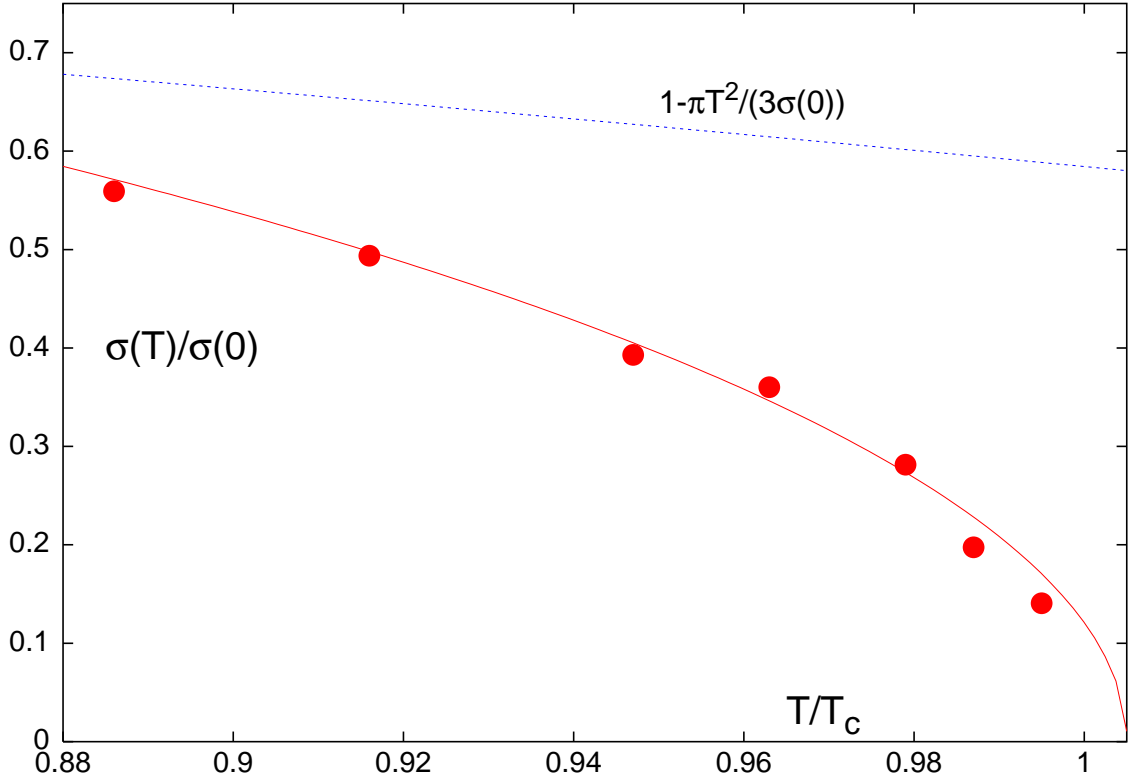,width=154mm}
\caption{The string tension as obtained from fits with
         eq.~\refe{gerrit2}, normalized to its zero temperature
         value. The line is the result of a fit to this ratio
         with the string model motivated ansatz eq.~\refe{mod_alva}.
         The data is compared with the lowest order temperature
         effect on the linear part of the potential,
         eq.~\refe{gerrit}, shown as the dotted line.
         }
\label{fig:olaf}
\end{figure}

Because of these findings, we fix this coefficient to 1
in the following. The resulting string tension,
normalized to its zero temperature value is shown in
Figure~\ref{fig:olaf}. The temperature dependence
compares well with the (modified) prediction \cite{Alvarez}
of the Nambu-Goto model, eq.~\refe{eq:alvarez},
\beqn
\frac{\sigma(T)}{\sigma(0)} = a\sqrt{1-b\frac{T^2}{T_c^2}}
\label{mod_alva}
\eqn
Recall that the string model
prediction assumes a second order transition with a
continuous vanishing of the string tension at the critical
temperature. The deconfinement transition in pure
$SU(3)$ Yang-Mills theory, however, is known to be of first order.
Thus, a discontinuity at the critical temperature is expected.
To account for this,
the coefficients $a$ and $b$ in eq.~\refe{mod_alva}
are allowed to deviate from unity.
In fact, the fit to the data, shown as the line in figure~\ref{fig:olaf},
results in the values $a=1.21(5)$ and $b=0.990(5)$.
This leads to a non-vanishing string tension at the
critical temperature of
\beqn
\frac{\sigma(T_c)}{\sigma(0)} = 0.121(35)
\eqn
This number can be converted into a value for the (physical) mass gap
at the transition point, $m_{\rm phys}(T_c)/T_c = \sigma(T_c)/T_c^2
= 0.30(9)$.
This is a bit below but not incompatible with earlier
results of dedicated analyses of the order of the deconfinement transition,
$m_{\rm phys}(T_c)/T_c = 0.4 - 0.8$, as summarized in \cite{Ukawa}.

Finally, we compared
the string tension $\sigma(T)$ defined in eq.~\refe{gerrit2}
with the leading behaviour
$\sigma(0)-\pi T^2/3$ as given in eq.~\refe{gerrit}.
This is shown as the dotted line in Figure~\ref{fig:olaf}.
Similarly to Figure~\ref{fig:gao} the
comparison fails, reflecting
that non-leading terms contribute substantially.

\section{Results above $T_c$}
\label{above}

Above the critical temperature we have normalized the Polyakov
loop correlations to their cluster value
\beqn
V(|\vec{R}|,T) = - T \ln \frac{\langle L(0) L^\dagger(\vec{R})
    \rangle}{|\langle L(0) \rangle |^2}
\label{corr_above}
\eqn
to eliminate the self-energy contributions.
In principle, the correlation function itself is periodic in $R$.
Alternatively, one can fit the potential, eq.~\refe{corr_above},
with a periodic ansatz, $V(R) \rightarrow V(R) + V(N_\sigma a -R)$.
The second contribution turns out to be very small
at the distances fitted and both procedures lead to
the same results for the fit parameters.

\begin{figure}[htb]
\epsfig{file=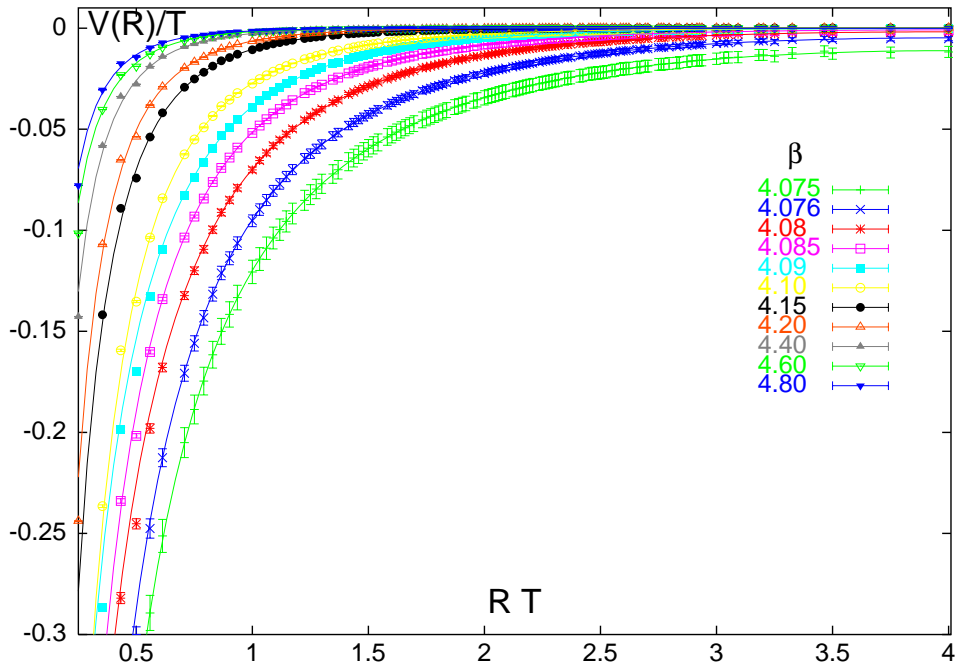,width=134mm}
\caption{The potentials $V(R,T)$ for the $\beta$ values above $\beta_c$
         analyzed on the $N_\tau=4$ lattices.
         The Polyakov loop correlations have been normalized to
         their cluster value. Potentials and distances are
         given in units of the temperature.
         }
\label{fig:nt4_high}
\end{figure}

In the following we first concentrate on data set (I) which
has somewhat better statistics and which, more important, covers
the explored range of distances more densely,
see Figure~\ref{fig:nt4_high}.
As has been explained in section~\ref{theory},
we fit the potentials above $T_c$ with
the generalized screening ansatz, eq.~\refe{free_d},
where the exponent $d$ of
the Coulomb-like part is treated as a free parameter.
It turns out that the value of the
exponent and the value of the screening mass $\mu$
are strongly correlated. In particular at the higher
temperatures it is difficult to obtain fit
results which are stable under the variation of the minimum distance
included in the fit. These fluctuations have been taken into account
in our estimates of the error bars.

At the highest temperatures analyzed we observed that
at large quark separations the Polyakov-loop
correlation decreases below the cluster value.
In \cite{Gale} it was argued that finite momentum
contributions to the vacuum polarization tensor can give
rise to a modified screening function which undershoots
the exponential Debye decay at intermediate distances
and approaches the infinite distance limit from below.
Despite the high, yet limited precision of our data we are
not in the position, however, to confirm this suggestion. Instead,
we have taken an operational approach and have added an overall
constant to our fit ansatz.

\begin{figure}[htb]
\epsfig{file=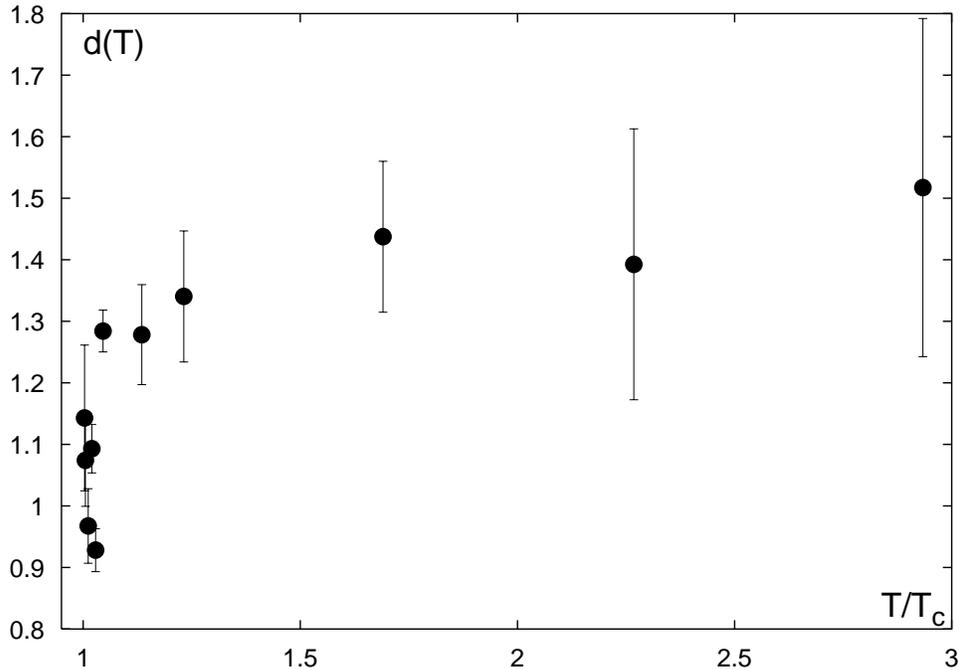,width=134mm}
\caption{Fit results for the exponent $d$ of the
         Coulomb-like part of the potential above $T_c$,
         eq.~\refe{free_d},
         as a function of the temperature.
         }
\label{fig:freed}
\end{figure}

In Figure~\ref{fig:freed} we summarize the results for
the exponent. At temperatures very close to $T_c$, the exponent
$d$ is compatible with $1$. When the temperature is
increased slightly, $d$ starts rising to about $1.4$ for temperatures
up to $2 T_c$. Between 2 and 3 times $T_c$, the exponent centers
around 1.5, although the error bars tend to become
rather large.
A value of 2 as predicted by perturbation theory
seems to be ruled out, however, in the investigated temperature range.

\begin{figure}[htb]
\epsfig{file=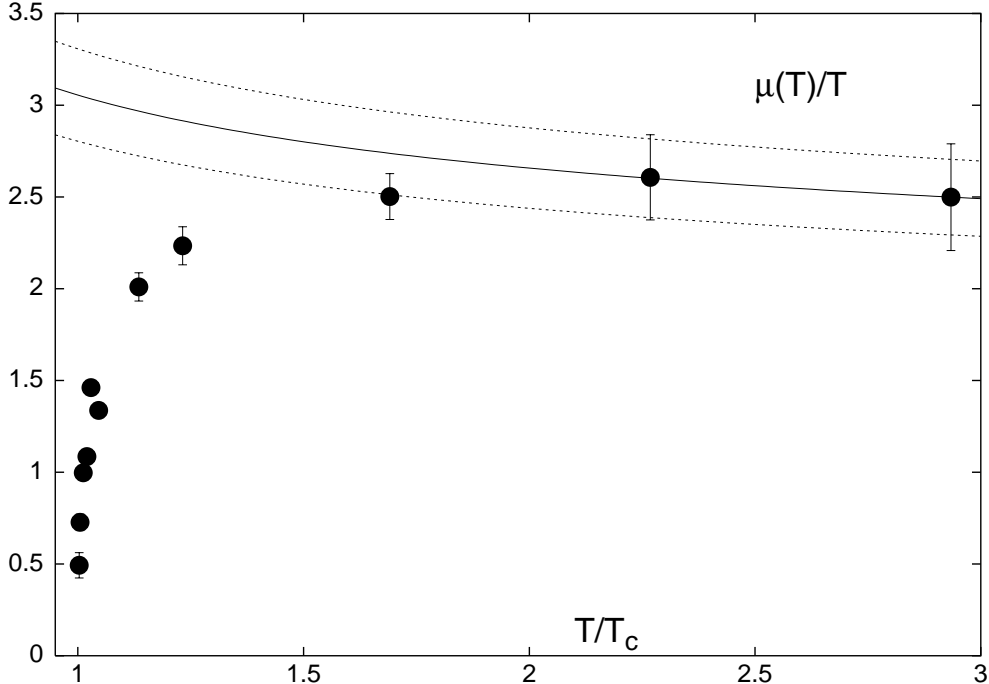,width=134mm}
\caption{Fit results for the screening mass $\mu(T)$,
         eq.~\refe{free_d},
         as a function of the temperature.
         The lines denote the comparison with the perturbative
         prediction, $\mu(T) = A \ast m_e^{(0)}(T)$,
         see text.
         }
\label{fig:mu}
\end{figure}

The results for the screening mass $\mu(T)$ obtained
from the same fits with eq.~\refe{free_d} are shown in
figure~\ref{fig:mu}. The screening mass turns out to be small
but finite just
above $T_c$ and rises rapidly when the temperature is increased.
It reaches a value of about $2.5 T$ at temperatures around
$1.5 T_c$ and seems to stabilize there also.
Figure~\ref{fig:mu} also includes a comparison with
lowest order perturbation theory, $\mu(T) = A \ast m_e^{(0)}(T)$
with $m_e^{(0)}(T)$ as given in eq.~\refe{pertmu}.
For the temperature dependent renormalized coupling $g^2(T)$
the two-loop formula
\beqn
g^{-2}(T) = 2 b_0 \ln\left(\frac{2 \pi T}{\Lambda_{\overline{MS}}}\right)
          + \frac{b_1}{b_0} \ln\left(\, 2 \ln \left(
            \frac{2 \pi T}{\Lambda_{\overline{MS}}}\right) \right)
\eqn
was used,
where $T_c/\Lambda_{\overline{MS}}=1.14(4)$ \cite{Beinlich,Bali}
and the lattice scale was set by the lowest Matsubara frequency
$2 \pi T$.
Perturbation theory predicts the factor $A$ to be 2.
Indeed, adjusting to the data points at the two highest
temperatures, $T > 2 T_c$, leads to a value of
$A = 1.82 \pm 0.15$ which is close to the prediction.
However, in view of the results for the exponent $d$ we regard this
as an accidental coincidence. This is further supported by
analyses \cite{Rank} in colour $SU(2)$ where the electric gluon mass
was obtained from gluon propagators and from the singlet potential
$V_1$, see eq~\refe{singlet_pot}. Here it was found that the observed
mass follows a behaviour $m_e(T) \simeq 1.6 m_e^{(0)}(T)$.
If this result could be transferred to the case of $SU(3)$,
and \cite{Attig} provides some early evidence for this,
we would have
$\mu(T)\simeq 1 \ast m_e(T)$ contrary to the perturbative
value of 2.

\begin{figure}[htb]
\epsfig{file=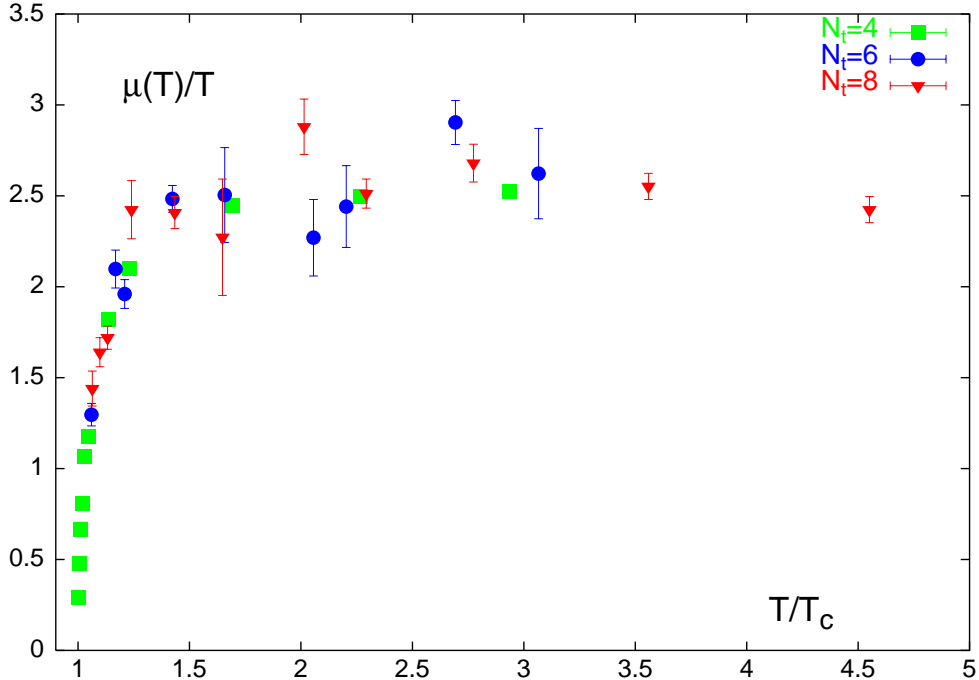,width=134mm}
\caption{The screening mass $\mu$ from fits with
         eq.~\refe{free_d} with a fixed value of $d=1.5$ for the
         exponent. The results from all three different
         lattice sizes are
         drawn as a function of the temperature.
         }
\label{fig:mufixed}
\end{figure}

The potentials above $T_c$ from data set (II) are very
similar to the ones already discussed.
Fits with eq.~\refe{free_d}
with a free exponent do work and return parameter values in the
same ballpark as in case I. However, because of the much smaller
number of distances probed in this set the fit results are not as
reliable as in case I. Therefore we have chosen to carry out
fits with eq.~\refe{free_d} but with $d$ kept fixed. For comparison,
also data set (I) has been treated this way.

The general feature of these fits is
that increasing $d$ from 1.0 to 2.0 leads to decreasing numbers
for the screening mass. For instance, at $3 T_c$ we obtain
$\mu/T \approx 3$ for $d=1.0$ whereas with $d=2.0$ the result
for the screening mass is $\mu/T \approx 2$. Similar shifts
occur at all temperatures. The quality of the fits, however,
is not always the same. Typically, at temperatures close to $T_c$
fits with $d=2$ return unacceptable $\Chi^2$ values while for
$T \geq 2 T_c$ the $\Chi^2$ values are equally good for all
values of $d$ and cannot be used
to distinguish between the various exponent values
anymore. This observation fits nicely into the picture as
shown in Figure~\ref{fig:freed}.

As an example for the temperature
dependence of the screening mass at fixed $d$,
in Figure~\ref{fig:mufixed}
we show our results at $d=1.5$ for all three
different lattice sizes. Recall that a value of $d \approx 1.5$
was favored at all temperatures $T \gsim 1.2 T_c$ of data set (I).
The general behavior
is similar to that shown in Figure~\ref{fig:mu}: the screening
mass is small close to $T_c$ and starts to rise quickly.
It reaches a kind of plateau with a value of $\mu /T \approx 2.5$
for temperatures between roughly 1.5 and 3 $T_c$. For temperatures
beyond $3 T_c$ the $N_\tau = 8$ data  may indicate a slow
decrease with rising temperature. The main conclusion to be drawn
from Figure~\ref{fig:mufixed} is that the results
from the different lattices, i.e. at different lattice spacings
are in agreement with each other within the error bars.
Thus, in the investigated temperature range colour screening
effects are not yet properly described by simple perturbative
predictions.

\section{Conclusion}
\label{conclusion}

In this paper we have analyzed the heavy quark potential
at finite temperatures in the range $0.8 T_c$ up to about
$4 T_c$ in $SU(3)$ Yang-Mills theory.
We have done so on lattices with 3 different
temporal extents and found results consistent with each other.
Moreover, the
standard Wilson action as well as a tree-level improved
Symanzik action were used. Again, consistency was observed.
This indicates
that finite lattice spacing artefacts are not futilizing
the analysis.

The potentials at temperatures below the critical temperature
of the deconfinement transition are well parametrized by
formulae which have been derived within string models.
In particular, the presence of a logarithmic
term with the predicted strength could be established.
However, the obtained string tension shows a substantial temperature
dependence which is not in accord with the leading
string model result. Instead, we find a decrease of the
string tension which is compatible with being proportional
to $(T_c-bT)^{1/2}$
in the critical region below $T_c$.
At the critical temperature the string
tension retains a finite value of $\sigma(T_c)/\sigma(0) =
0.121(35)$, consistent with a first order transition.

Above the deconfinement transition the potentials show a
screened power-like beha\-viour. By comparing the data
with perturbative predictions we can further strengthen
earlier claims that these predictions do not properly
describe the potentials up to temperatures of few times the
critical one. In particular, it can be excluded that the exchange
of two gluons with an effective chromo-electric mass is the
dominant
screening mechanism.
Judging from the exponent of the $1/R$ term in
the potential, at temperatures close to $T_c$ it seems
that the complex interactions close to the phase transition
arrange themselves in such a way as to be effectively describable
by some kind of one-gluon exchange.
At temperatures of about 1.5 to 3 times $T_c$ we observe a behaviour which
could be interpreted as a mixture of one- and two-gluon exchange.
The resulting screening mass scales with the temperature,
$\mu(T) \simeq 2.5 \, T$, a perturbative decrease due to the
temperature-dependent renormalized coupling $g(T)$ is not really seen.
Thus, it is very likely that non-perturbative phenomena and higher
order perturbative contributions are needed to explain the
observed screening behaviour in the investigated
temperature range.

\medskip
\noindent
{\bf Acknowledgements:}

\noindent
This work was supported by the TMR network ERBFMRX-CT-970122,
the DFG grant Ka 1198/4-1 and partly by the
``Nederlandse Organisatie voor Wetenschappelijk Onderzoek'' (NWO)
via a research program of the ``Stichting voor Fundamenteel
Onderzoek der Materie'' (FOM).
The numerical work has been carried out
on Quadrics QH2 and QH1 computers at the University of Bielefeld
which in part have been funded by the DFG under grant Pe 340/6-1.
F.K. acknowledges support through the visitor program of the
Center for Computational Physics at the University of Tsukuba and
thanks the CCP for the kind hospitality extended to him.
E.L. thanks the NIKHEF for the kind hospitality and J. Koch
for critical comments on the manuscript.

\newpage

\section*{Appendix}

\begin{table}[h]
\begin{center}
\begin{tabular}{|ll|c|l|}
\hline
$N_\tau$ & $\beta$ & $T/T_c$ & $\sigma a^2$ \\
\hline
4        & 3.95  & 0.804 & 0.2256(6) \\
         & 4.00  & 0.886 & 0.1783(4) \\
         & 4.02  & 0.916 & 0.1581(2) \\
         & 4.04  & 0.947 & 0.1339(4) \\
         & 4.05  & 0.963 & 0.1263(3) \\
         & 4.06  & 0.979 & 0.1113(8) \\
         & 4.065 & 0.987 & 0.0970(8) \\
         & 4.07  & 0.995 & 0.0882(15) \\
\hline
6        & 5.80  & 0.832 & 0.0917(10) \\
         & 5.85  & 0.919 & 0.0634(23) \\
         & 5.87  & 0.956 & 0.0526(8) \\
         & 5.89  & 0.993 & 0.0203(13) \\
         &       &       & 0.0400(20) \\
\hline
8        & 6.00  & 0.904 & 0.0386(6) \\
         & 6.04  & 0.967 & 0.0204(5) \\
         &       &       & 0.0300(15) \\
         & 6.06  & 0.9985& 0.0078(7) \\
         &       &       & 0.0195(15) \\
\hline
\end{tabular}
\end{center}
\caption{Results for the string tension in lattice units from fits
         with eq.~\refe{Gao}. In those cases where two lines
         are printed for a given $\beta$ the upper line refers
         to averages over the whole sample while the lower line
         takes into account only configurations which are in the
         confined phase.
         }
\label{sigma}
\end{table}

\begin{table}
\begin{center}
\begin{tabular}{|ll|c|l|}
\hline
$N_\tau$ & $\beta$ & $T/T_c$ & $\sigma a^2$ \\
\hline
4        & 3.95  & 0.804 & 0.1592(8) \\
         & 4.00  & 0.886 & 0.1125(4) \\
         & 4.02  & 0.916 & 0.0925(3) \\
         & 4.04  & 0.947 & 0.0686(3) \\
         & 4.05  & 0.963 & 0.0607(3) \\
         & 4.06  & 0.979 & 0.0458(6) \\
         & 4.065 & 0.987 & 0.0316(9) \\
         & 4.07  & 0.995 & 0.0221(11) \\
\hline
\end{tabular}
\end{center}
\caption{Results for the string tension from fits with
         eq.~\refe{gerrit}.
         }
\label{sigma2}
\end{table}

\end{document}